# The Nonlocal Nature of Natural Deduction & the Introduction of Assumptions in Derivations

**Khashayar Irani**
**Birkbeck College, University of London**
**k.irani@mathematicallogic.com**
**October 2026**

**Abstract**

This concise manuscript investigates the nonlocal nature of natural deduction, particularly the possibility of introducing assumptions at intermediate stages of a derivation after those stages have already been passed.[1] By examining the structural correspondence between natural deduction and sequent calculus, we demonstrate how the structural rule of left-weakening (LW) permits the enlargement of an antecedent after its corresponding succedent has been derived, without modifying the original inference.[2] Through a number of symmetrical derivations, we establish the proof-theoretic relationship between the introduction of assumptions in natural deduction and antecedent expansion in sequent calculus. On this basis, we formulate the principle of 'Nonlocal Assumption Introduction' (NAI), according to which assumptions may be introduced at any intermediate stage prior to the final conclusion, provided that the corresponding structural expansion is proof-theoretically valid.[3] We further

---

[1] This study, together with a number of other investigations that will be published either prior to or alongside it, proposes several structural analyses concerning sequent calculus and natural deduction that we believe have been overlooked in previous and contemporary works in proof theory. The central thesis of the present paper was originally inspired by certain proof-theoretic analyses developed in Khashayar Irani (2026), which prompted us to examine more systematically the structural properties governing the introduction of assumptions at intermediate stages of natural deduction derivations. In the future, we aim to produce a comprehensive manuscript that brings together these results and presents them as fundamental structural principles of these proof systems.

[2] Note that the principal concern of this proof-theoretic analysis is with those natural deduction calculi that are extracted from or correspond to structural sequent calculus systems satisfying the structural principles of weakening, contraction and cut. Consequently, natural deduction constructions based on substructural proof systems are not the focus of our enquiry.

[3] We maintain, of course, that NAI has always been an inherent structural feature of natural deduction systems grounded in structural sequent calculi, and that our modest contribution consists in explicitly identifying and formulating this principle. Indeed, as the subsequent analysis will demonstrate, the operation characterised by NAI is so deeply embedded in the structural organisation of natural deduction that its presence becomes virtually impossible to overlook once the relevant proof-theoretic correspondence is made explicit. A further area for future research concerns the possibility of formulating a principle of 'Nonlocal Arbitrary Conclusion Introduction' (NACI), grounded in the classical structural principle of right-weakening (RW). Unlike LW, which permits the addition of an arbitrary antecedent formula, RW permits the enlargement of the succedent of an already derived sequent, namely $\Gamma \Rightarrow \Delta \vdash \Gamma \Rightarrow \Delta, A$. This suggests investigating whether a corresponding operation in multiconclusion natural deduction could permit the addition of a further arbitrary conclusion after

specify the conditions governing this operation, including the preservation of inferential validity, assumption dependencies, and the scope and discharge requirements of the resulting derivation.[4]

Natural deduction is conventionally represented as a progressively constructed proof system in which premises and assumptions are introduced, inference rules are applied, intermediate formulae are derived, and the derivation advances towards its final conclusion.[5] This graphical presentation suggests a chronological conception of proof construction, according to which an intermediate stage, once completed and subsequently passed, may appear inaccessible to the later introduction of assumptions. The present investigation challenges this conception by distinguishing the graphical order of a derivation from its underlying proof-theoretic dependency structure. Although the construction of a proof necessarily follows a particular order, this order does not independently determine whether the structural context associated with an already established derivational state may subsequently be enlarged. Consider an intermediate formula B derived under an assumption context $\Gamma$, represented in sequent calculus by $\Gamma \Rightarrow B$. Suppose that an additional assumption [A] is introduced after B has already been derived.[6] Since [A] was not employed in the original derivation of B, the relevant question is not whether [A] contributed to that derivation, but whether the context $\Gamma$ may subsequently be expanded to include [A] without altering the derivational validity of B.

---

an intermediate derivational stage has been completed, without modifying the original inference, while preserving inferential validity and the relevant structural requirements.

[4] We define inferential validity as the formal legitimacy of an inference within a specified proof system, such as David Hilbert's axiomatic system or Gerhard Gentzen's natural deduction and sequent calculus. An inference is valid when its conclusion is derived from its premises through the correct application of the inference rules governing that system. Inferential validity is therefore determined by the logical and structural conditions governing derivations rather than by the psychological processes of reasoning. Unlike semantic validity, which concerns truth preservation across interpretations, inferential validity concerns the formal correctness of derivations. For a brief description of a proof system and the three major proof systems in proof theory, see Irani (2025) Page: 789.

[5] Apart from the thesis advanced in this paper, namely the technique of NAI, our method of presenting natural deduction proof trees follows the Prawitz-Gentzen tradition. For a detailed account of this tradition, see Gentzen, cited in M. E. Szabo (1969), Dag Prawitz (1965 & 1971), and Sara Negri & Jan von Plato (2001 & 2011). Furthermore, the proof-theoretic role of assumptions in natural deduction has been examined from perspectives other than their introduction and structural dependency. For example, Edward Hermann Haeusler (2015) investigates the relationship between assumption repetition and proof complexity in purely implicational minimal and classical propositional calculi. Through a recursively defined family of formulae, Haeusler demonstrates that certain tautologies in minimal logic require at least 2^n occurrences of the same assumption in every normal natural deduction proof, whereas classical reasoning permits substantially shorter derivations. His findings also have implications for cut-free sequent calculus and automated proof procedures. Although Haeusler's investigation concerns the number of assumption occurrences required within derivations, rather than their introduction at previously completed stages, it illustrates the broader proof-theoretic significance of assumptions.

[6] As is our established practice, throughout the natural deduction constructions as well as in the body of the text, we enclose assumed formulae in [ ].

Sequent calculus provides a formal mechanism for precisely this operation through the structural rule of LW:[7]

$$\frac{\Gamma \Rightarrow \Delta}{A,\Gamma \Rightarrow \Delta}\,\text{LW}$$

The significance of LW lies in the fact that its premise Γ ⇒ Δ is already derived independently of [A], whereas its conclusion extends the antecedent by introducing [A] while preserving the succedent Δ. Consequently, when LW is applied to the sequent Γ ⇒ B, the resulting sequent A,Γ ⇒ B represents an enlargement of the antecedent rather than a modification of the inference through which B was originally derived. This establishes an important distinction between the inferential grounds of a formula and the structural context within which that formula may subsequently occur. In particular, an assumption introduced after an intermediate formula has been derived need not be regarded as having participated in its original derivation purely because its corresponding formula subsequently belongs to the enlarged antecedent. The nonlocal character of natural deduction proposed in this study rests upon this distinction, since the chronological priority of an intermediate formula does not necessarily prevent the subsequent expansion of its associated assumption context. Nevertheless, such an expansion remains subject to the ordinary proof-theoretic requirements governing the validity of inference rules, the dependencies of formulae, and the scope and discharge of assumptions. Thus, the proposed nonlocality concerns the possibility of structurally legitimate assumption introduction beyond the graphical stage at which an intermediate formula was originally derived, rather than an unrestricted freedom to modify completed derivations.

The structural significance of LW can be demonstrated through two symmetrical derivations in which an additional antecedent formula is introduced only after the intermediate sequent A,¬A ⇒ ⊥ has been derived. These constructions provide a formal illustration of the relationship between antecedent expansion in sequent calculus and the introduction of assumptions in natural deduction, without requiring any alteration to the logical inferences preceding LW. Consider the following sequent calculus derivations and their corresponding natural deduction counterparts:[8]

---

[7] For a comprehensive study of various sequent calculus systems and related topics, including structural rules such as LW, see Negri & von Plato (2001 & 2011) and Anne Sjerp Troelstra & Helmut Schwichtenberg (2000).

[8] These sequent calculus and natural deduction derivations were originally presented in Irani (2026) to demonstrate that the principles of R¬ and ¬I permit the derivation of arbitrary conclusions through the introduction and subsequent discharge of arbitrary assumptions, thus providing a proof-theoretically motivated alternative to the conventional employment of RW, L⊥ and ⊥E within classical proof systems.

$$\dfrac{\dfrac{\dfrac{\mathbf{A \Rightarrow A}}{\mathbf{A,\neg A \Rightarrow \bot}}\,\mathbf{L\neg}}{\mathbf{\neg C,A,\neg A \Rightarrow \bot}}\,\mathbf{LW}}{\mathbf{A,\neg A \Rightarrow C}}\,\mathbf{R\neg} \qquad \dfrac{\dfrac{\dfrac{\mathbf{A \Rightarrow A}}{\mathbf{A,\neg A \Rightarrow \bot}}\,\mathbf{L\neg}}{\mathbf{C,A,\neg A \Rightarrow \bot}}\,\mathbf{LW}}{\mathbf{A,\neg A \Rightarrow \neg C}}\,\mathbf{R\neg}$$

$$\dfrac{\dfrac{[\mathbf{A}] \qquad [\mathbf{\neg A}] \quad [\mathbf{\neg C}]}{\bot}\,\mathbf{\neg E}}{\mathbf{C}}\,\mathbf{\neg I} \qquad \dfrac{\dfrac{[\mathbf{A}] \qquad [\mathbf{\neg A}] \quad [\mathbf{C}]}{\bot}\,\mathbf{\neg E}}{\mathbf{\neg C}}\,\mathbf{\neg I}$$

In both sequent derivations, L¬ is applied to the reflexive sequent A ⇒ A to derive A,¬A ⇒ ⊥, independently of the additional formulae [¬C] and [C]. Subsequently, LW enlarges the antecedent, deriving ¬C,A,¬A ⇒ ⊥ and C,A,¬A ⇒ ⊥, respectively, while preserving the original derivation of the contradiction. The applications of R¬ then complete the corresponding sequent derivations. Similarly, the natural deduction counterparts introduce [¬C] and [C] without requiring either assumption for the original derivation of ⊥ from [A] and [¬A]. These constructions demonstrate that the subsequent inclusion of an assumption within a derivational context does not entail its participation in the inference through which an earlier formula was derived. Consequently, the structural expansion permitted by LW provides a formal representation of assumption introduction after an intermediate derivational stage has been completed, without modifying the inference responsible for that stage.[9] The correspondence between natural deduction assumptions and sequent antecedents can be illustrated more directly through the following elementary derivation:

$$\dfrac{[\mathbf{A}] \qquad [\mathbf{\neg A}]}{\bot}\,\mathbf{\neg E}$$

Here, the assumptions [A] and [¬A] correspond to the antecedent formulae A and ¬A in the sequent A,¬A ⇒ ⊥. This correspondence establishes the representational basis for examining how an already derived intermediate formula may subsequently be associated with an enlarged assumption context in natural deduction.

[9] A proof-theoretically beneficial employment of NAI in deriving classical theorems, such as Peirce’s law, can be found in Irani (2026), where we derive ((A → B) → A) → A without employing ⊥E. Unlike the conventional derivation, which obtains the arbitrary formula B through ⊥E, our construction introduces the additional assumption [¬B] after deriving ⊥ from [A] and [¬A]. This application of NAI permits the derivation of B through the discharge of [¬B] by ¬I. Thus, NAI, together with ¬I, provides a structurally legitimate alternative to ⊥E.

The preceding analysis can be extended to derivations in which an intermediate formula is derived through an application of an inference rule before the antecedent is expanded by LW. This provides a more substantial illustration of NAI, since the inferential origin of the intermediate formula can be identified independently of the subsequent structural operation. Consider the inferred conditional A → B, from which B is derivable together with [A] through the natural deduction rule →E:

$$\dfrac{\mathbf{A \to B} \qquad \mathbf{[A]}}{\mathbf{B}}\ \mathbf{\to E}$$

The corresponding sequent calculus derivation employs L→:

$$\dfrac{A \Rightarrow A \qquad\qquad B \Rightarrow B}{\mathbf{(A \to B),A \Rightarrow B}}\ \mathbf{L\to}$$

to derive (A → B),A ⇒ B from the reflexive sequents A ⇒ A and B ⇒ B. Once this sequent has been derived, LW permits the introduction of an additional antecedent formula, namely [¬C] or [C], without requiring another application of L→. The following symmetrical constructions:

$$\dfrac{\dfrac{\dfrac{\mathbf{A \Rightarrow A} \qquad \mathbf{B \Rightarrow B}}{\mathbf{(A \to B),A \Rightarrow B}}\ \mathbf{L\to}}{\mathbf{\neg C,(A \to B),A \Rightarrow B}}\ \mathbf{LW}}{\mathbf{(A \to B),A \Rightarrow B,C}}\ \mathbf{R\neg}
\qquad\qquad
\dfrac{\dfrac{\dfrac{\mathbf{A \Rightarrow A} \qquad \mathbf{B \Rightarrow B}}{\mathbf{(A \to B),A \Rightarrow B}}\ \mathbf{L\to}}{\mathbf{C,(A \to B),A \Rightarrow B}}\ \mathbf{LW}}{\mathbf{(A \to B),A \Rightarrow B,\neg C}}\ \mathbf{R\neg}$$

$$\dfrac{\dfrac{\mathbf{A \to B} \qquad \mathbf{[A]} \qquad \mathbf{[\neg C]}}{\mathbf{B}}\ \mathbf{\to E}}{\mathbf{B} \qquad\quad \mathbf{C}}\ \mathbf{\neg I}
\qquad\qquad
\dfrac{\dfrac{\mathbf{A \to B} \qquad \mathbf{[A]} \qquad \mathbf{[C]}}{\mathbf{B}}\ \mathbf{\to E}}{\mathbf{B} \qquad\quad \mathbf{\neg C}}\ \mathbf{\neg I}$$

provide a further demonstration of the relationship between the derivation of an intermediate formula and the subsequent expansion of its assumption context.[10] In both sequent calculus constructions, L→ derives (A → B),A ⇒ B from the reflexive sequents A ⇒ A and B ⇒ B, corresponding to the derivation of B from A → B and [A] through →E in natural deduction. Subsequently, LW introduces [¬C] or [C] into the antecedent, deriving ¬C,(A → B),A ⇒ B and C,(A → B),A ⇒ B, respectively, without altering the inference responsible for B. The applications of R¬ then derive (A → B),A ⇒ B,C and (A → B),A ⇒ B,¬C, while the corresponding natural deduction constructions associate [¬C] and [C] with the applications of ¬I. Despite their different polarities, both constructions exhibit the same structural operation, namely the enlargement of an already derived antecedent through LW. Importantly, the additional assumption belongs to the expanded dependency context without having participated in the original derivation of B. Thus, the chronological introduction of an assumption must be distinguished from its structural inclusion within the context of an intermediate formula. The fact that B has already been derived through →E does not independently prohibit the subsequent introduction of [¬C] or [C], provided that the resulting derivation remains formally valid. These constructions therefore illustrate how an assumption context may be enlarged after its associated intermediate formula has been derived, without reconstructing the original inference. They also identify the proof-theoretic requirements that must govern such an operation, particularly the preservation of inferential validity, assumption dependencies, and the scope and discharge conditions of the additional assumption.

[10] Like the preceding derivations, these sequent calculus and natural deduction constructions were originally presented in Irani (2026) to demonstrate that the principles of R¬ and ¬I permit the derivation of arbitrary conclusions through the discharge of arbitrary assumptions, even within contradiction-free multiconclusion derivations. Their original purpose was to establish that the proof-theoretic applicability of R¬ and ¬I extends beyond traditional contradiction-based reductio constructions, without requiring the employment of RW, L⊥ or ⊥E.

The structural correspondence demonstrated by the preceding derivations permits the explicit formulation of a principle governing the introduction of assumptions at intermediate stages of natural deduction, which we designated as the NAI principle. As emphasised earlier, this rule is not intended to introduce a new inference rule or modify the established architecture of natural deduction. Rather, it identifies a structural possibility already inherent in natural deduction calculi grounded in structural sequent calculus systems admitting weakening, contraction and cut. Its formulation rests upon the distinction between the inferential grounds upon which an intermediate formula is originally derived and the subsequent expansion of the assumption context associated with that formula. Nevertheless, the admissibility of such an expansion is subject to specific proof-theoretic conditions. First, the intermediate formula must have been legitimately derived under its original assumption context. Second, the introduction of an additional assumption must be warranted by a corresponding structural operation, particularly LW, without altering the inference responsible for the original derivation. Third, all subsequent applications of inference rules must remain formally legitimate, preserving the dependencies, scope and discharge conditions governing the assumptions involved. To formulate NAI more precisely, let Π be a natural deduction derivation containing an intermediate formula B derived under an assumption context Γ, represented in the corresponding sequent calculus by Γ ⇒ B. Suppose that the graphical construction of Π has advanced beyond the original occurrence of B and that an additional assumption [A] is subsequently introduced into the context associated with that intermediate stage. Since LW warrants the transition from Γ ⇒ B ⊢ A,Γ ⇒ B, the introduction of [A] may be understood as the corresponding structural enlargement of the natural deduction assumption context, provided that the stated conditions are satisfied. Accordingly, NAI states that an assumption [A] may be introduced at any intermediate stage of a natural deduction derivation prior to its final conclusion, even after the graphical construction has passed that stage, whenever the corresponding antecedent expansion is warranted by LW and the resulting derivation preserves inferential validity and the relevant assumption dependencies. The expression 'at any stage' must therefore be understood in a structural rather than an unrestricted graphical sense. In particular, NAI does not authorise the retrospective treatment of an additional assumption as a premise employed in deriving an earlier formula, nor does it permit violations of the conditions governing assumption discharge. Its applicability depends upon the legitimate correspondence between the natural deduction dependency context and the antecedent of an already derived sequent. In this manner, NAI characterises the admissible introduction of assumptions into previously established derivational contexts while preserving the original inferences and the structural requirements governing their subsequent employment.

To conclude, the principal objective of this research has been to examine the nonlocal nature of natural deduction and, more specifically, to establish the proof-theoretic legitimacy of introducing assumptions at intermediate stages of a derivation after those stages have already been passed. We began by distinguishing the graphical order in which a natural deduction proof is constructed from the structural dependencies governing its intermediate formulae. This distinction provided the foundation for examining the relationship between natural deduction and sequent calculus, particularly the principle of LW. We demonstrated that LW permits the enlargement of an antecedent after its corresponding succedent has been derived, without altering the inference responsible for that derivation. Subsequently, through symmetrical sequent calculus and natural deduction constructions, we illustrated how additional assumptions may be introduced into an established derivational context without having participated in the original inference. The first group of derivations employed contradiction and R¬ and ¬I, whereas the second employed implication and the corresponding rules L→ and →E before introducing additional assumptions through LW. These constructions established the distinction between the inferential origin of an intermediate formula and the subsequent structural enlargement of its assumption context. Importantly, the examples also demonstrated that this structural operation is applicable to both singleconclusion and multiconclusion proof systems, provided that the corresponding inference rules and structural requirements are satisfied. Thus, the analysis supplied the formal motivation for identifying NAI as a structural feature of natural deduction rather than an independent inferential innovation.

The findings of this paper also open a number of avenues for further proof-theoretic research. Of particular interest is the development of a more general account of the relationship between NAI and the structural principles governing different families of natural deduction and sequent calculus systems. A second area concerns the implications of NAI for normalisation, particularly whether its application can be systematically accommodated within normalisation procedures while preserving the relevant proof-theoretic properties of derivations. Another promising direction concerns the formulation of a corresponding principle of 'Nonlocal Arbitrary Conclusion Introduction' (NACI), grounded in the structural rule of RW. Just as NAI concerns the enlargement of an antecedent, NACI would concern the possibility of introducing an additional arbitrary conclusion into an already established derivational context without modifying the inference responsible for the original conclusion. Such a principle could be investigated as a fundamental structural property of multiconclusion natural deduction, particularly in relation to the correspondence between the introduction of additional conclusions and succedent expansion in sequent calculus. Further research may also examine the applicability and limitations of NAI and NACI within substructural proof systems, thus clarifying the extent to which these principles depend upon the availability of particular structural rules. These questions provide a foundation for developing a more comprehensive account of the structural organisation of natural deduction and its relationship with sequent calculus.